\documentclass{jfm}
\usepackage{graphicx}
\usepackage{natbib}
\usepackage{bm}

\ifCUPmtlplainloaded \else
  \checkfont{eurm10}
  \iffontfound
    \IfFileExists{upmath.sty}
      {\typeout{^^JFound AMS Euler Roman fonts on the system,
                   using the 'upmath' package.^^J}%
       \usepackage{upmath}}
      {\typeout{^^JFound AMS Euler Roman fonts on the system, but you
                   dont seem to have the}%
       \typeout{'upmath' package installed. JFM.cls can take advantage
                 of these fonts,^^Jif you use 'upmath' package.^^J}%
      }
  \else
  \fi
\fi

\ifCUPmtlplainloaded \else
  \checkfont{msam10}
  \iffontfound
    \IfFileExists{amssymb.sty}
      {\typeout{^^JFound AMS Symbol fonts on the system, using the
                'amssymb' package.^^J}%
       \usepackage{amssymb}%

      }{}
  \fi
\fi

\ifCUPmtlplainloaded \else
  \IfFileExists{amsbsy.sty}
    {\typeout{^^JFound the 'amsbsy' package on the system, using it.^^J}%
     \usepackage{amsbsy}}
    {\providecommand\boldsymbol[1]{\mbox{\boldmath $##1$}}}
\fi

\newsavebox{\astrutbox}
\sbox{\astrutbox}{\rule[-5pt]{0pt}{20pt}}

\def\R{{\bf R}}
\def\Z{{\bf Z}}
\def\Sy{{\bf S}}

\title[Transition in pipe flow]{Transition in pipe flow: the saddle structure on the
boundary of turbulence}

\author[Y. Duguet, A.P. Willis and R.R. Kerswell]%
{Y. \ns  D\ls U\ls G\ls U\ls E\ls T,$^1$%
\thanks{Present address : Linn\'e Flow Centre, KTH Mechanics, SE-100 44 Stockholm, Sweden.} \ns A.\ls P.\ns W\ls I\ls L\ls L\ls I\ls S $^1$
\thanks{Present address : Laboratoire d'Hydrodynamique, Ecole Polytechnique, 91128 Palaiseau, France}\break
\and R.\ls R.\ns K\ls E\ls R\ls S\ls W\ls E\ls L\ls L$^1$
}

\affiliation{$^1$ School of Mathematics,
   University of Bristol,
   BS8 1TW Bristol, United Kingdom\\[\affilskip]
}
\pubyear{2007}
\volume{1}
\pagerange{1--2}
\date{November 13th 2007 and in revised form May 27th 2008}
\usepackage{epsfig}
\usepackage{latexsym}

\begin{document}

\maketitle

\begin{abstract}

  The laminar-turbulent boundary $\Sigma$ is the set separating
  initial conditions which relaminarise uneventfully from those which
  become turbulent. Phase space trajectories on this
  hypersurface in cylindrical pipe flow look to be chaotic and show
  recurring evidence of coherent structures. A general numerical
  technique is developed for recognising approaches to these
  structures and then for identifying the exact coherent solutions
  themselves. Numerical evidence is presented which suggests that
  trajectories on $\Sigma$ are organised around only a few travelling
  waves and their heteroclinic connections. If the flow is suitably
  constrained to a subspace with a discrete rotational symmetry, it is
  possible to find locally-attracting travelling waves embedded within
  $\Sigma$. Four new types of travelling waves were found using this
  approach.

\end{abstract}

%\pacs{47.20.Ft,47.27.Cn,47.60.+i}

%*******INTRODUCTION********************************************
\section{Introduction \label{sec:intro}}

Transition to turbulence in cylindrical pipe flow is governed by one
single dimensionless parameter, the Reynolds number $Re:=UD/\nu$ where
$U$ is the mean flow speed along the pipe, $D$ the pipe diameter and
$\nu$ the kinematic viscosity of the fluid (Reynolds 1883). Despite
the simplicity of the set-up, the reason for transition remains
obscure due to the linear stability of the laminar Hagen-Poiseuille
flow (Hagen 1839, Poiseuille 1840) and the sensitivity of the process
to the exact shape and amplitude of disturbances present. In most
experiments, transition is observed at $Re \sim 2000$ (e.g. Wygnanski
\& Champagne 1973) but can be triggered as low as $Re=1750$ (Peixinho
\& Mullin 2006) or delayed to $Re=100,000$ in very carefully
controlled experiments (Pfenniger 1961). Until recently, the only firm
theoretical result was the energy stability bound of $Re=81.49$
(Joseph \& Carmi, 1969) below which {\it all} disturbances are
guaranteed to decay monotonically.  This is, however, more than an
order of magnitude below the observed value for transition.

An important step forward in understanding the transition process was
the discovery of disconnected solutions to the Navier-Stokes equations
in a cylindrical pipe (Faisst \& Eckhardt 2003, Wedin \& Kerswell
2004, Kerswell 2005, Pringle \& Kerswell 2007). These exact solutions
are travelling waves (TWs) which appear through saddle-node
bifurcations as in other shear flows (Nagata 1990, Waleffe
1997,1998,2001,2003). All these solutions are linearly unstable though
they have a very low-dimensional unstable manifold.  There is much
interest in these solutions, as very similar structures have been
observed transiently in experiments (Hof et al.  2004, 2005) and
direct numerical simulations (Kerswell \& Tutty 2007, Schneider
\emph{et al.} 2007a, Willis \& Kerswell 2008a).  These states can
generally be divided into `upper-branch' and `lower-branch' TWs, based
on whether they have high or low wall shear stress. Lower branch
solutions are believed to sit on a hypersurface that divides phase
space into two regions: one where initial points lead directly to the
laminar state, the other where initial conditions lead to turbulent
episodes (Kawahara 2005, Wang {\it et. al.} 2007, Kerswell \& Tutty
2007, Viswanath 2008). The simple translational behaviour of these
travelling wave solutions is inherent to the method used to find them,
and undoubtedly masks an even larger variety of more complex exact
solutions.

The boundary between laminar and turbulent trajectories - labelled
$\Sigma$ hereafter - is formally a separatrix if the turbulent state
is an attractor. At low $Re$, however, turbulence may ultimately
decay after a long transient in which case the laminar state is the
unique global attractor. Then the boundary $\Sigma$ is generalised
to the dividing set in phase space between trajectories which
smoothly relaminarise and those which do undergo a turbulent
evolution. $\Sigma$ is thought to be of codimension 1 in phase space
but one can \emph{a priori} not exclude a more complex fractal
structure, as suggested by the dependence of lifetime on initial
conditions in simulations of pipe flow (Faisst \& Eckhardt 2004).
Trajectories which start in $\Sigma$ - hereafter called the `edge'
following Skufca et al. (2006) - stay in $\Sigma$ for later times by
definition and hence the long time dynamics are of obvious interest.
The long time behaviour on $\Sigma$ has already been found to be a
periodic solution in plane Poiseuille flow in a pioneering study by
Toh \& Itano (1999, see also Itano \& Toh 2001). Skufca et al.
(2006) studied a 9-dimensional model of plane Couette flow (PCF) to
reveal an attracting periodic orbit at low $Re$ and an apparently
chaotic state at higher $Re$. However, recent fully-resolved
simulations have shown that the asymptotic behaviour is an
attracting TW in PCF (Schneider  et al. 2008, Viswanath 2008) and a
chaotic attractor in a short cylindrical pipe of length $L=5\,D$
(Schneider et al. 2007b). Interestingly, this seemingly chaotic end
state looks to be centered around an `asymmetric' TW
(Pringle \& Kerswell 2007, Mellibovsky \& Meseguer 2007).\\

The purpose of this paper is to explore the dividing hypersurface
$\Sigma$ in pipe flow with the following objectives:
\begin{enumerate}
\item to establish that the dynamics restricted to this
laminar-turbulent boundary explores many different saddle points
embedded in it;
\item
to find evidence for heteroclinic or `relative' homoclinic
connections between these saddle points;
\item to explore $\Sigma$ restricted by a discrete rotational
symmetry in order to ascertain whether the limiting behaviour
remains chaotic or can be a simple attractor;
\item to develop a practical and general way to find TWs and periodic
orbits without
detailed knowledge of their spatial structure.\\
\end{enumerate}

The works cited above employ a reduced computational domain for
their calculations, imposing short-wavelength periodicity in one
direction (along the pipe, Schneider et al. 2007b) or two (in the
spanwise and streamwise directions in plane Couette flow, Schneider
et al. 2008, Viswanath 2008). This is very much in the spirit of the
Minimal Flow Unit pioneered by Jimenez \& Moin (1991) and used with
considerable success to identify key mechanisms underpinning
turbulence (Hamilton et al. 1995). In this paper, we also adopt this
same approach by concentrating on pipes up to $5\,D$ long.  The
ensuing reduction in the degrees of freedom of the flow allows a
much more detailed exploration of the flow dynamics albeit
restricted to a subspace of strict spatial periodicity. This is a
necessary preliminary to motivate a more carefully focussed study in
much longer pipes which can capture spatially-localised turbulent
structures (e.g. Priymak \& Miyazaki 2004, Willis \& Kerswell 2007,
2008a) but at considerable numerical cost still. Flow structures
found in the short pipes considered here also exist, of course, in a
longer pipe but will be more unstable due to the greater degrees of
freedom present there. It is also worth remarking that a recent
study (Nikitin 2007) has shown how pipe flow is slow to lose its
spatial periodicity when this is restriction is lifted.

This paper is organised as follows. Section 2 discusses the
formulation and numerical methods used to simulate the flow in a pipe
and to extract exact recurrent flow solutions. Section 3 presents the
results obtained using these in 7 subsections.  Subsection 3.1 recalls
the method used to follow trajectories on $\Sigma$ and confirms that
the limiting behaviour looks chaotic (Schneider et al. 2007b).
Sections 3.2 and 3.3 discuss how near-recurrent states are identified
and using a Newton-Krylov algorithm demonstrate that a number of
unstable travelling waves are approached. Subsection 3.4 offers
evidence for the existence of a `relative' homoclinic connection
between a travelling wave and the same wave rotated. Subsection 3.5
looks for recurrent flow structures in $\Sigma$ restricted by a
discrete rotational symmetry and identifies new exact travelling wave
solutions. Subsection 3.6 shows that within this subspace the limiting
state of $\Sigma$ is a simple attractor. Subsection 3.7 confirms that
all the TWs found in the earlier subsections are indeed embedded in
the laminar-turbulent boundary. The paper ends with a discussion in
section 4.

\section{Numerical Procedure \label{sec:computing}}

\subsection{Governing Equations \label{sec:eqns}}

We consider the incompressible flow of Newtonian fluid in a
cylindrical pipe and adopt the usual set of cylindrical coordinates
$(s,\theta,z)$ and velocity components ${\bm u}= u \hat{{\bm s}} + v
\hat{{\bm \theta}} + w \hat{{\bm z}}$.  The domain considered here is
$(s,\theta,z) \in [0\!:\!1]\times[0\!:\!2\pi]\times[0\!:\!L]$, where
$L=2\pi/\alpha$ is the length of the pipe and lengths are in units of
radii ($D/2$).  The flow is described by the incompressible
three-dimensional Navier-Stokes equations
\begin{eqnarray}
\frac{\partial {\bm u}}{\partial t} + \left({\bm u}\cdot {\bm
\nabla}\right){\bm u} &=& - {\bm \nabla}p +
\frac{1}{Re}\nabla^{2}{\bm u},\\
{\bm \nabla} \cdot {\bm u} &=& 0.
\end{eqnarray}
The flow is driven by a constant mass-flux condition, as in recent
experiments (e.g. Peixinho \& Mullin 2006). The boundary conditions
are periodicity across the pipe length ${\bm u}(s,\theta,z)={\bm
u}(s,\theta,z+L)$ and no-slip on the walls ${\bm u}(1,\theta,z)={\bm
0}$.

\subsection{Time-stepping code \label{sec:code}}

The basic tool for the numerical determination of exact recurrent
states is the accurate time-stepping code used by Willis \& Kerswell
(2007,2008a). The velocity field is derived from two scalar
potentials $\Psi$ and $\Phi$:
\begin{eqnarray}
{\bm u}= {\bm \nabla} \times (\Psi {\bm \hat{z}}) + {\bm \nabla}
\times {\bm \nabla} \times (\Phi {\bm \hat{z}})
\end{eqnarray}
and the incompressible Navier-Stokes equations are re-written using
the formulation introduced by Marqu\'es (1990). The two scalar
potentials are discretised using high-order finite differences in
the radial direction $s$ and spectral Fourier expansions in the
azimuthal direction $\theta$ and axial direction $z$. For example,
the decomposition of the scalar potential $\Phi$ at a radial
location $s_j$, $(j=1,...,N)$, reads
\begin{eqnarray}
\Phi(s_j,\theta,z,t;\alpha,m_0)= \sum_{k=-K}^{K}
\sum_{m=-M}^{M}\Phi_{jkm}(t)e^{i(m_{0}m\theta +\alpha k z)}
\label{representation}
\end{eqnarray}
The positive integer $m_0$ refers to the discrete rotational symmetry
\begin{eqnarray}
 \R_{m_{0}}:(u,v,w,p)(s,\theta,z) \rightarrow (u,v,w,p)(s,\theta+\frac{2\pi}{m_{0}},z)
 \end{eqnarray}
of the flow (trivially $m_0=1$ means no rotational symmetry is
imposed). The resolution of a given calculation is described by a
vector $(N,M,K)$. It is adjusted until the energy in the spectral
coefficients drops by at least 5 and usually 6 decades from lowest
to highest-order modes. The time stepping is 2nd order accurate with
$\Delta t$ updated using an adaptive method based on a CFL
condition.

The number of (real) degrees of freedom, which defines the dimension
of phase-space, is $O(8MNK)$. The set of all complex coefficients
${\bm X}=\{\Phi_{jkm},\Psi_{jkm}\}$ defines our phase-space with its
usual Euclidean norm $|{\bm X}|=\sqrt{{\bm X} \cdot {\bm X}}$.
Symmetries exhibited by TWs such as the shift-and-reflect symmetry
\begin{eqnarray}
\Sy:(u,v,w,p)(s,\theta,z)\rightarrow
(u,-v,w,p)(s,-\theta,z+\frac{\pi}{\alpha}),
\end{eqnarray}
and the mirror symmetry (e.g. Pringle \& Kerswell 2007, Pringle et
al. 2008)
\begin{eqnarray}
\Z:(u,v,w,p)(s,\theta,z)\rightarrow (u,-v,w,p)(s,-\theta,z),
\end{eqnarray}
are not imposed in the simulations.

\subsection{The Newton-Krylov method \label{sec:newton}}

The spectral expansion defined above converts the Navier-Stokes
equations into an autonomous dynamical system of the form
\begin{eqnarray}
\frac{d{\bm X}}{dt}={\bm F}({\bm X}). \label{dxdt}
\end{eqnarray}
Travelling wave solutions are `relative equilibrium' solutions of the
Navier-Stokes, steady in an appropriate Galilean frame. It is
numerically convenient to seek and interpret them as a special case of
``relative periodic orbit" (RPO) (Viswanath 2007, 2008). We hence
developed an algorithm to look for RPOs defined as zeros of the
functional
\begin{equation}
g=|{\bm X}(T)^{-\Delta z,-\Delta \theta}-{\bm X}(0)|^{2}.
\end{equation}
Here, ${\bm X}(T)$ is the point at time $T$ on the trajectory
starting at time $t=0$ from the point ${\bm X}(0)$. ${\bm
X}^{-\Delta z,-\Delta \theta}$ is the point in phase space
corresponding to the state ${\bm X}$ shifted back in space by the
distance $\Delta z$ in the axial direction and by the angle $\Delta
\theta$ in the azimuthal direction. A shift by $(\Delta z,
\Delta \theta)$ corresponds in phase-space to the transformation
\begin{eqnarray}
(\Psi_{jkm},\Phi_{jkm})e^{-i(mm_{0}\Delta \theta + \alpha k \Delta z)}
\rightarrow
(\Psi_{jkm},\Phi_{jkm}).
\end{eqnarray}
A zero of $g$ corresponds to a flow repeating itself exactly after a
time T, but at a different location defined by $\Delta z$ and
$\Delta \theta$. A travelling wave solution is a RPO where there is
a degeneracy between the shifts  $\Delta z$ and $\Delta \theta$, and
the period $T$. For the case of a TW propagating axially with speed
$c$, $\Delta z=cT$, where $T$ is an apparent period. To remove this
degeneracy, we impose $\Delta z=L$ unless otherwise specified. As
the majority of known TWs do not rotate, $\Delta \theta=0$ is also
assumed (those that do rotate do so rather slowly, Pringle \&
Kerswell, 2007). Starting from a good initial guess for ${\bm X}(0)$
and an estimate of the period $T$ (to be discussed in Section
\ref{sec:rec}), we minimise the residual $g$ using a Newton-Krylov
algorithm, based on a GMRES algorithm (Saad \& Schultz, 1986). The
size of the dynamical system (typically $10^5$ degrees of freedom)
necessitates the use of a matrix-free formulation.  The use of an
\emph{inexact} Krylov solver also leads to a significant gain in
computation time (we used the choice 2 in Eisenstat \& Walker, 1995).  Moreover, we embed
the Newton solver into a more globally convergent strategy in order
to improve likeliness of convergence, using a double dogleg step
technique (Dennis \& Schnabel, 1995, Brown \& Saad 1990).  For the
special case of TWs with a short period, we assume full convergence
when the normalised residual $\sqrt{g}/|{\bm X}(0)|$ is less than
$O(10^{-10})$.

\subsection{Stability of Travelling Waves \label{sec:stability}}

Once a travelling wave solution is known, along with its axial
propagation speed $c$, it can be expressed as a steady solution in the
frame moving at speed $c$. In this Galilean frame, the stability of
the solutions can be studied numerically using an eigenvalue solver
based on an Arnoldi algorithm.  This yields the leading eigenvalues
whose real part, when positive, indicates the growth rate of
infinitesimal perturbations to the exact solution in the moving frame.

\section{Results \label{sec:results}}

We present 4 `edge' calculations, each motivated by a different
question.  In the first (see \S \ref{sec:edge}-3), the edge in a
pipe length $L \approx 5\,D$ ($\alpha=0.625$) is calculated starting
from a typical turbulent initial condition in order to investigate
whether there are coherent structures buried in it. We are able to
confirm the presence of what looks to be a chaotic attractor as
found recently by Schneider et. al. (2007b). In the second edge
calculation (\S \ref{sec:heteroc}) we look for numerical evidence of
heteroclinic connections between saddle points by using a perturbed
TW as an initial condition. In the third (\S \ref{sec:m_02}) we
impose the discrete rotational symmetry $\R_2$ on the flow to
exploit the saddle structure of the subset of $\Sigma$ to reveal new
TWs which possess $\R_2$-symmetry. In the fourth (\S
\ref{sec:attractor}) we look for evidence of multiple attracting TWs
to demonstrate that under a rotational symmetry constraint the
large-time dynamics need not be chaotic.

\subsection{Calculating Edge Trajectories \label{sec:edge}}

In this subsection we choose a pipe with $\alpha=0.625$
($L=5.0265\,D$) and $Re=2875$, set $m_0=1$ so that there is no
restriction on the rotational symmetry, and take a numerical
resolution of $(30,15,15)$. For this value of $Re$, turbulence, when
triggered, is sustained over the typical simulation time which is
much longer that the time taken to relaminarise. In order to
constrain a numerical trajectory to stay on the edge surface,
$\Sigma$, we use a shooting method analogous to that first used in
plane Couette flow by Toh \& Itano (1999). We first produce a long
turbulent trajectory and pick a state ${\bm u}^{*}$ of relatively
low three-dimensional disturbance energy $E_{3d}(t)$ (defined as the
kinetic energy in the streamwise-dependent Fourier modes $k \neq
0$). Initial states are then chosen from the line,
\begin{eqnarray}
{\bm u}_{\beta} := \langle {\,\bm u}^* \rangle + \beta({\bm u}^* -
\langle{\bm u}^{*} \,\rangle) ,
\label{ubeta}
\end{eqnarray}
parameterised by the real number $\beta$ and $\langle \quad \rangle:=
{\alpha \over 2\pi} \int_{0}^{\frac{2\pi}{\alpha}}(..)dz$ is an axial
average. The initial condition ${\bm u}_{\beta=0}$ is two-dimensional
and hence does not trigger turbulence, whereas ${\bm u}_{\beta=1}$ is
the original turbulent state.  This $\beta$-interval is then
repeatedly halved with, at each stage, the new interval
selected such that an initial condition corresponding to $\beta$ at
the lower limit relaminarises, whereas that at the upper limit leads to
large turbulent values of $E_{3d}$. $\beta$ is refined close to
machine accuracy using double precision arithmetic, forcing the energy
of the trajectory to stay at an intermediate level corresponding to
the laminar-turbulent boundary for typical times of $O(200\,D/U)$. At
the values of $Re$ used here, the energy of this boundary is clearly
distinct from those associated with turbulence and thus makes
determination of $\beta$ unambiguous. Once machine precision has been
reached, the process is restarted from nearby states towards the end
of the trajectory originating from neighbouring $\beta$ values.

\begin{figure}
\includegraphics{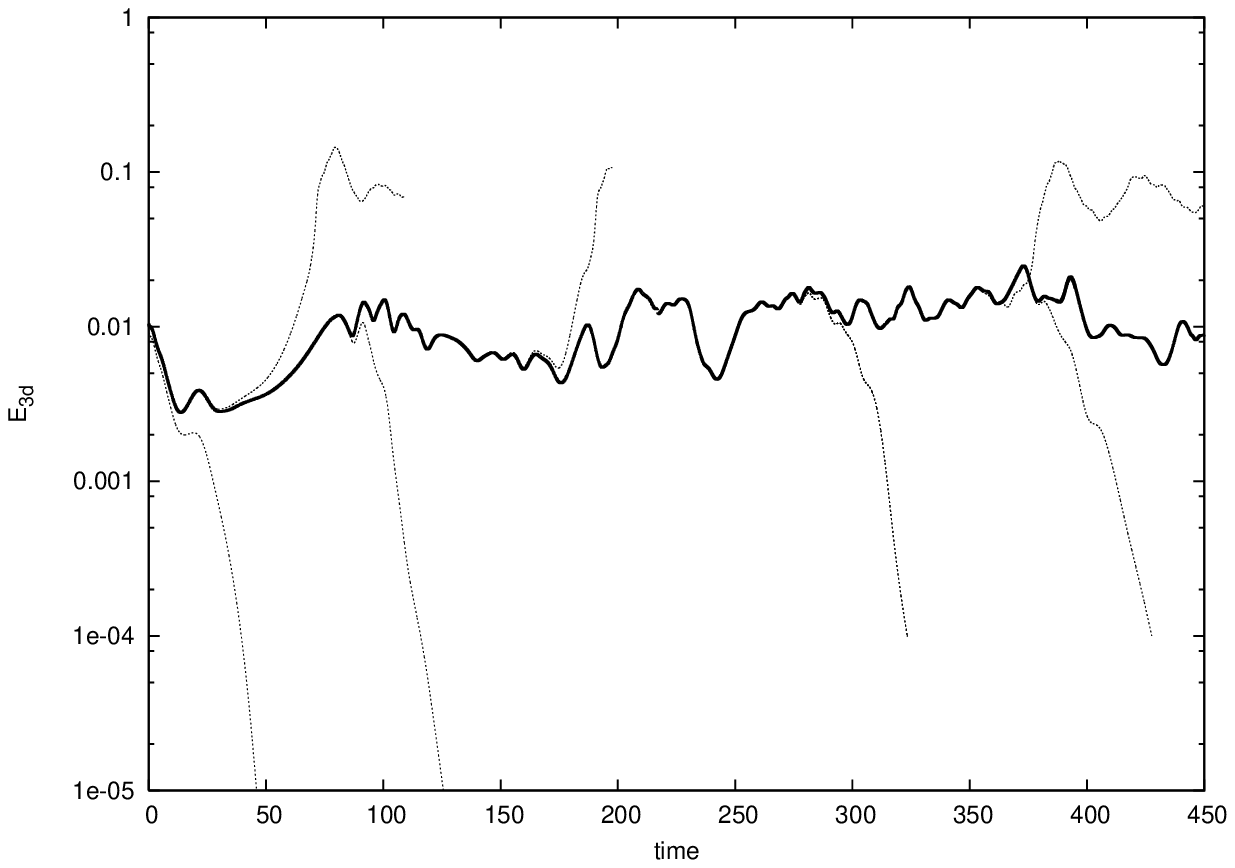}
\caption{Energy contained in the axially-dependent modes for
$(\alpha,Re,m_0)=(0.625,2875,1)$. The thick line indicates the edge
trajectory and the thinner lines nearby trajectories which either
relaminarise (energy decreases) or becomes turbulent (energy
increases to a higher level). Time is in units of $D/U$.}
\label{E1}
\end{figure}

The full trajectory, of duration $O(500\, D/U)$, does not show
any sign of convergence towards a simple state but rather displays
erratic and unpredictable dynamics, as found by Schneider et al (2007b).
We deliberately avoid the word `turbulent' as this refers to larger
values of $E_{3d}$. Figure \ref{E1} shows the time evolution of the
$E_{3d}(t)$.

\subsection{Near-Recurrences on an Edge Trajectory \label{sec:rec}}

\begin{figure}
\includegraphics[width=14cm]{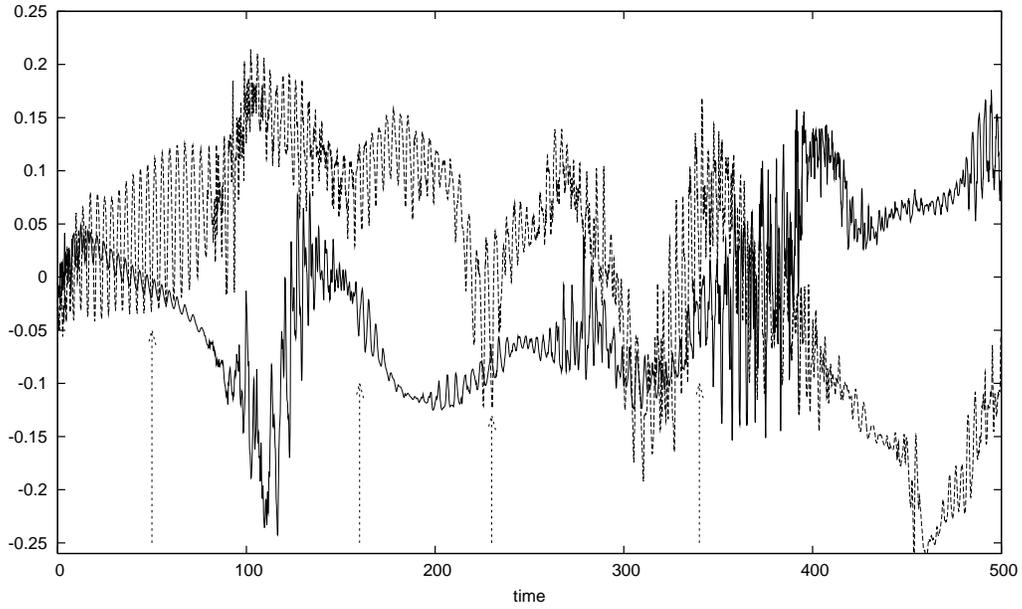}
\caption{Axial perturbation velocity signal at two different
locations in the pipe: ($s=0.66,\theta=0,z=0$) and
($s=0.76,\theta=0,z=0.5\,D$). The signal is taken from the edge
trajectory in Section \ref{sec:edge}. Dotted arrows indicate where
the velocity signal shows approximate recurrences.}
 \label{velz}
\end{figure}

\begin{figure}
\includegraphics{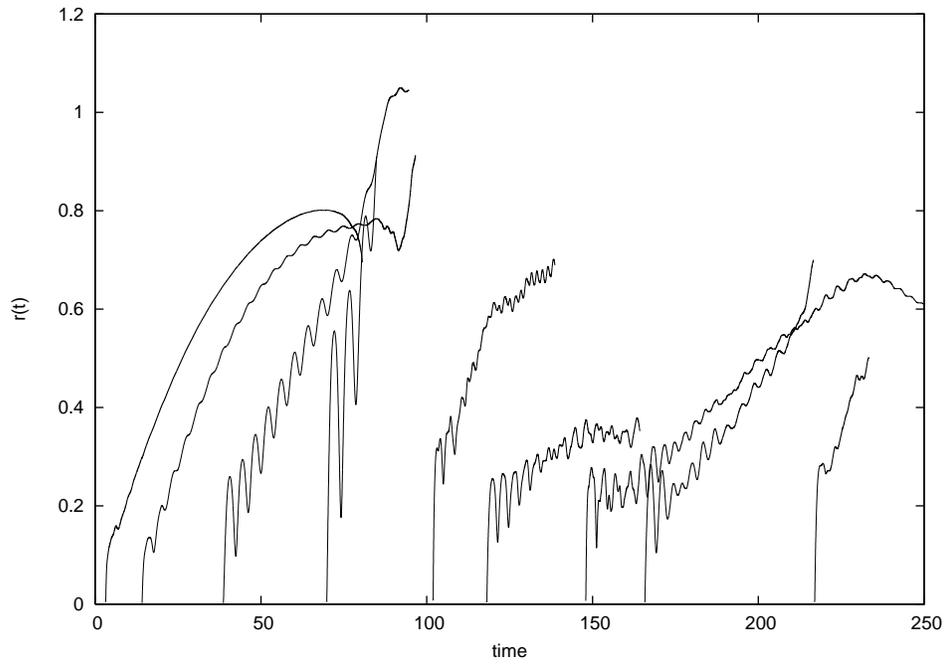}
\caption{Typical profiles of the residual function $r(t)$ starting
from snapshots of the edge trajectory of Section \ref{sec:edge}. The
subscript $i$ has been suppressed (time in units of $D/U$).}
\label{res1}
\end{figure}

Despite the lack of regularity of the energy signals, inspection of
all velocity components at several random locations in the pipe
indicate clearly that the flow on $\Sigma$ is sometimes nearly
periodic in time on short intervals.  For the case
$(\alpha,Re,m_0)=(0.625,2875,1)$, it is possible by eye to identify
several temporal windows where all velocity components approximately
oscillate with a given frequency (see Figure \ref{velz}). Within these windows the dynamics on the edge appear to be temporally
correlated with a short period.  A large number of snapshots were
taken at times $t_i, i=1,2,3...$ across the whole trajectory. To
investigate the possibility of recurrence in the flow, each snapshot
state ${\bm X}(t_i)$ was used as an initial condition for
time-stepping.  The normalised distance in phase space between this
initial point and its evolution in time was then examined for local
minima. Specifically, we define the scalar residual function
\begin{eqnarray}
r_{i}(t>t_{i}):= \frac{|{\bm X}^{-\Delta z, -\Delta \theta}(t)-{\bm
X}(t_i)|}{|{\bm X}(t_i)|}
\end{eqnarray}
where $\Delta z$ is a distance by which the state ${\bm X}(t)$ is
shifted back in $z$ for comparison. We chose $\Delta z=L$ so that a
value of $r_i(t)=0$ for some $t>t_i$ means that the flow in the pipe
is \emph{exactly} the flow at time $t_i$. In this case, ${\bm X}(t_i)$
lies on a periodic orbit of the system with the time interval $t-t_i$
being a multiple of the period (for $\Delta z \neq 0$, a vanishing
residual would indicate a \emph{relative} periodic orbit). For the
ease of calculation, most of the time we chose to neglect the
possibility of recurrence occurring after a shift in the azimuthal
direction $\theta$, and therefore $\Delta \theta =0$ was imposed.
Typical plots of $r_i(t)$ starting from different values of $t_i$ are
displayed in Figure \ref{res1}. Every function $r_i$ attains a
smallest local minimum which is defined as
\begin{eqnarray}
r_{min}(t_i) := \min_{t>t_i}\{\, r_{i}(t): \,  \frac{\partial
r_{i}}{\partial t}=0 \,\}.
\end{eqnarray}
%
%The scalar function $r_{min}$, which can be determined for any given
%initial condition ${\bm X}(t_i)$, turns out to be
%the central quantity of this paper.

\begin{figure}
\includegraphics{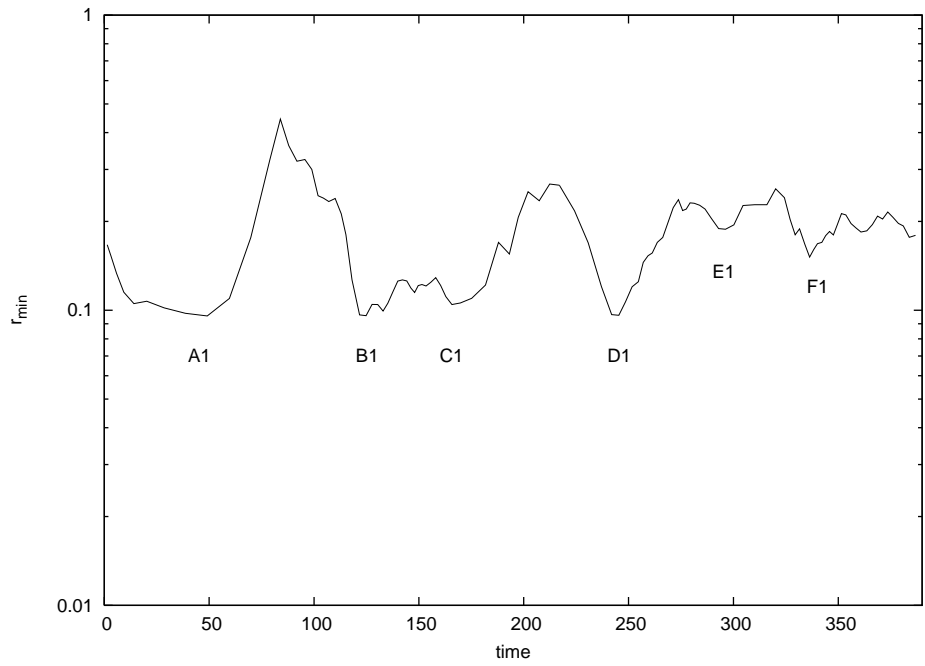}
\caption{Recurrence signal for the edge trajectory of Section
\ref{sec:edge} (for definition of $r_{min}$ see text) against time
in $D/U$. $(\alpha,Re,m_0)=(0.625,2875,1)$.}
\label{rmin1}
\end{figure}

Figure \ref{rmin1} shows how $r_{min}$ varies with the starting point.
Phases for which $r_{min}$ is small (from experience below $0.1$) may
be interpreted as approximate approaches to periodic orbits of the
system by the edge trajectory. The alternating pattern of maxima and
minima in $r_{min}$ is then the signature of a repellor - a set of
states which attract trajectories before ultimately repelling them
away. For hyperbolic dynamical systems, trajectories are attracted
towards one of these states along their stable manifold and ejected
away along their unstable manifold (see the sketch in Figure
\ref{phasespace}). For the parameters here, 6 dips in the
$r_{min}$-profile are suggestive of 6 approaches, denoted respectively
by $A1, B1, C1, D1, E1, F1$.  Parts of the trajectory linking one
state to the next (e.g. $A1 \rightarrow B1$ or $D1 \rightarrow E1$)
may be located in the vicinity of heteroclinic connections between the
two states, should such a connection exist, or a relative homoclinic
connection for two successive states which are the same but for a
shift.  The notion of `vicinity' depends on the choice of the norm
in phase space. Here, a pragmatic approach is adopted: we use the
expression `${\bm X}$ is close to a periodic orbit $Y$' to mean that
the Newton-Krylov algorithm converges to the periodic orbit $Y$
starting with the initial guess ${\bm X}$.

\subsection{Exact Coherent Structures in $\Sigma$ \label{sec:exact}}

We now analyse the neighbourhood of the points ${\bm X}(t_i)$ yielding
a low value of $r_{min}$ looking for exact recurrent states for which
$r_{min}$ exactly vanishes. Starting from such initial guesses, the
Newton-Krylov algorithm defined in \S \ref{sec:newton} was used to
further minimise $r_{min}$.  Of the six different starting guesses A to F
introduced in the previous subsection, excellent convergence
was obtained in the cases $A1$, $B1$, $D1$ and  $F1$, with $r_{min}$ being
reduced down to $O(10^{-11})$. The converged states are labelled
respectively $A1\_0.625$, $B1\_0.625$, $D1\_0.625$ and $F1\_0.625$ in
order to distinguish them from the starting points $A1,...,F1$, and to
indicate the parameter $\alpha$. The cases $C1$ and $E1$ displayed
only partial convergence to respectively $O(10^{-3})$ and $O(10^{-2})$
and it is not possible to say if there really are zeros of $r_{min}$
in this neighbourhood. Despite `globalisation' improvements to the
Newton-Krylov algorithm to improve convergence, sample starting
guesses away from the candidates $A,...,F$ failed to converge.  The
converged states $A1\_0.625$, $B1\_0.625$, $D1\_0.625$ and $F1\_0.625$
all correspond to travelling wave solutions. Close inspection of their
spatial structure and dynamics shows that they are, in fact, the {\em
  same} travelling wave solution modulo a shift in the azimuthal
direction: see Figure \ref{plots1}. This state corresponds exactly
to the `asymmetric' TW identified in Pringle \& Kerswell (2007)
whose $z$-averaged velocity field is strongly reminiscent of the
chaotic state calculated by Schneider et al. (2007b). This
resemblance has also been noted by Mellibovsky \& Meseguer (2007)
who used a different approach to infer the significance of this same
asymmetric TW. Interestingly, the present study at these parameter
values did not find evidence of states with  higher rotational
symmetry in spite of the fact that such TWs are known to be embedded
in the edge (see fig. 4 of Kerswell \& Tutty, 2007).

The search for recurrent states described above was initially
undertaken with $\Delta \theta=0$ disallowing azimuthal propagation of
the recurrent patterns. In the case $A1$, we experimented with
allowing $\Delta \theta$ to be updated by the Newton-Raphson scheme.
This involved solving an additional equation $\partial g / \partial
\Delta \theta=0$ (Viswanath 2007). Using Newton-Krylov with the same
starting guess $A1$ but allowing the shift $\Delta \theta$ to be
updated, produced convergence to another state $A'\_0.625$ which is
distinct from $A1\_0.625$. This state, despite a velocity profile very
analogous to that of $A1\_0.625$, rotates by an angle $\Delta \theta =
-0.32$ degrees after travelling one pipe length.  This situation is
reminiscent of the bifurcation diagram obtained by Pringle \& Kerswell
(2007) for the asymmetric TW (though here $Re$ is much higher): for a
given Reynolds number, a branch of solutions with helicity is
connected to the mirror-symmetric branch, and intersects the
non-helical subspace twice. The solutions at the crossing points are
non-helical but nevertheless possess a rotational propagation speed
$c_{\theta} \neq 0$, like our solution $A'\_0.625$. The fact that such
a solution has been found here could be explained by the fact that the
numerical code used does not allow helicity, hence the Newton-Krylov
algorithm has no choice but to look for the intersection of the
helical branch with a non-helical subspace. However, other attempts to
find exact recurrent patterns with rotation invariably converged back
to a non-rotating ($\Delta \theta=0$) TW.

\subsection{Search for Heteroclinic Connections \label{sec:heteroc}}

The results so far indicate the possible existence of heteroclinic
trajectories linking the exact states found. When the two
consecutively visited states are the same modulo a shift in the
azimuthal direction, ``relative" homoclinic connection is a more
appropriate term. In an attempt to find further evidence for such
connections, a series of initial conditions sampling the unstable
manifold of a TW were explored. The asymmetric travelling wave of
Pringle \& Kerswell (2007) was chosen (i.e. $A1\_0.625$) with the
parameters $(\alpha,Re,m_0)=(0.625,2875,1)$ and orientated so that
it was $\Sy$-symmetric about the $\theta=0$ plane. In the
corresponding $\Sy$-symmetric subspace, the solution has exactly two
unstable eigendirections, denoted by ${\bm e}_1$ and ${\bm e}_2$,
${\bm e}_1$ being the most unstable one (there are no unstable
directions in the dual space of $\Sy$-{\em anti}symmetric
perturbations). These two vectors were normalised to have unit
energy and an initial condition was defined as
\begin{eqnarray}
{\bm X}(0) = {\bm X}_{TW} + \epsilon \left( \cos{\phi}\,{\bm e}_1 +
\sin{\phi} \, {\bm e}_2 \right).
\end{eqnarray}
Here, $\epsilon$ is a positive parameter chosen small enough that
${\bm X}(0)$ is effectively in the unstable manifold of ${\bf
X}_{TW}$ ($\epsilon =10^{-2}$ suffices). The angle $\phi$ defines a
plane spanned by ${\bm e}_1$ and ${\bm e}_2$ and vanishes in the
direction of ${\bm e}_1$. For some values of the angle $\phi$, the
trajectory starting at ${\bf X}(0)$ returns to the laminar state
whereas for others, it becomes turbulent. Boundary values of $\phi$
between these two behaviours lead to intermediate dynamics
demonstrating that ${\bm X}_{TW}$ sits on the edge $\Sigma$. In all
cases the trajectory starts in the $\Sy$-symmetric subspace, but after
a finite time this symmetry is broken by numerical noise. A
bisection method is used to determine (to machine precision) an
angle $\phi$ for which the trajectory neither evolves to the laminar
or turbulent state (see Figure \ref{heteroplot} for a sketch of the
method). The energy contained in the axially-dependent modes is
displayed in Figure \ref{E4} and the residual function $r_{min}(t)$
in Figure \ref{rmin4}. These show that $r_{min}$ increases
exponentially but the energy changes little until $ t \sim 150 D/U$.
After this linear regime, $r_{min}$ continues to increase, then
drops to a local minimum value of $0.09$ at $t \sim 250 D/U$ before
increasing again. When the state at the local minimum (B4 in Figure
\ref{rmin4}) was used as a starting point for the Newton-Krylov
algorithm, $r_{min}$ smoothly decreased down to $O(10^{-11})$. The
exact solution found was the asymmetric travelling wave again but
now shifted by an angle of 51.56 degrees. On the basis that the
Newton-Krylov method usually needs to be in the vicinity of an exact
solution to converge to it, this suggests the existence of a
relative homoclinic connection near the computed trajectory.
Collecting firmer evidence is difficult unless the dynamics is
further restricted. Recently, significant progress has been made
along these lines in plane Couette flow by Gibson {it et al.} (2008)
who have demonstrated the existence of a heteroclinic connection
between two steady solutions on $\Sigma$ for a minimal flow unit.

\subsection{$\Sigma$ under $\R_2$-Symmetry \label{sec:m_02}}

The technique developed above to identify recurrent states in the
edge can also equally be applied to the dynamics within a symmetry
subspace. Restricting the flow to be $\R_m$-symmetric, for example,
improves the possibility of an edge trajectory coming close to some
of the $\R_m$-symmetric lower branch TWs found to be in the edge by
Kerswell \& Tutty (2007). There is also, of course, the possibility
of discovering unknown branches of TWs. Attention was restricted to
the $\R_2$-symmetric subspace by setting $m_0=2$ in the flow
representation (\ref{representation}).  At $Re=2875$, $\alpha=0.625$
and resolution $(30,15,15)$, an apparently chaotic edge trajectory was followed for $O(500 D/U)$
during which 5 different phases of relative recurrence - labelled
$A2,B2,C2,D2$ and $E2$ in chronological order - were detected.
Interestingly, this edge trajectory, although seemingly chaotic, is
much smoother than the corresponding (same $Re$ and $\alpha$) edge
trajectory for $m_0=1$: compare Figures \ref{E1} and \ref{E2}. Of
the 5 initial conditions for a Newton-Krylov search, $r_{min}$ only
converged to $O(10^{-11})$ for $D2$ and $E2$ ($r_{min}$ for $A2, B2$
and $C2$ stagnated at $O(10^{-3})$).

The exact recurrence states found, $D2\_0.625$ and $E2\_0.625$ (see
Figure \ref{plots2}), correspond to two new types of travelling wave
branches distinct from those originally reported in Faisst \&
Eckhardt (2003) and Wedin \& Kerswell (2004). $D2\_0.625$ is
especially interesting as it is the first TW found in pipe flow
which is {\em not} $\Sy$-symmetric. It does, however, possess a
mirror symmetry
\begin{equation}
\Z_\psi: \, (u,v,w,p)(s,\phi,z) \rightarrow
(u,-v,w,p)(s,2\psi-\phi,z)
\end{equation}
where $\psi \approx 45^o$ (or equivalently $\approx -45^o$ because of
the imposed $\R_2$-symmetry) for the snapshot shown in Figure
\ref{plots2}. In contrast, $E2\_0.625$ has three symmetries: two
unimposed - $\Sy$ and $\Z_{\pm \pi/4}$ (where $\phi=0$ defines the
plane of shift-\&-reflect symmetry) - and one imposed - $\R_2$.
$E2\_0.625$ is actually one member of a whole new class of {\em
  highly}\,-symmetric TWs which have subsequently been uncovered (the
$N$-class, see Pringle et al. 2008). These new waves are significant
because they appear much earlier in $Re$ than the originally
discovered TWs (Faisst \& Eckhardt 2003, Wedin \& Kerswell 2004) and
their upper branch solutions look more highly nonlinear.

\subsection{Local relative attractors within the $\R_2$-symmetric
subspace
\label{sec:attractor}}

In this subsection, the flow is again restricted to be
$\R_2$-symmetric but the pipe is halved in length ($\alpha=1.25$ so $L
\approx 2.5\,D$) and $Re$ reduced to $2400$ (resolution is
$(50,20,8)$). At these settings, Kerswell \& Tutty (2007) identified
a lower branch TW - $2b\_1.25$ in their nomenclature - which was on
the edge with only one unstable direction normal to it. This TW is
therefore a local attractor for trajectories confined to the edge. A
leading question is whether it is also a global attractor on the edge.\\

A starting point was randomly chosen along a turbulent trajectory,
and an edge trajectory  generated for $500\, D/U$ using the method
described above. The energy signal is displayed in Figure \ref{E3}
and the corresponding recurrence signal $r_{min}(t)$ is displayed in
Figure \ref{rmin3}. Two successive dips in $r_{min}$ can be seen at
times $50\, D/U$ and $90\, D/U$ later corresponding to inexact
recurrences labelled respectively $A3$ and $B3$. When used to
initialise a Newton-Krylov search, both led to a value of
$O(10^{-11})$ for $r_{min}$. The two converged exact states
correspond to one unique travelling wave solution, differing only by
a small shift in the azimuthal direction of less than 4 degrees. As
a result, we label only the first $A3\_1.25$: see Figure
\ref{plots2}. The TW $A3\_1.25$ has all the same symmetries as
$E2\_0.625$ but possesses a distinctly different structure
consisting of 4 low-speed streaks near the walls separated by 4
high-speed streaks. Four other high-speed streaks of lesser
amplitude and smaller size, and with a much more steady structure,
can be found closer to the pipe axis. This new TW thus has a richer
structure than the $\R_2$ branch of solutions found in Faisst \&
Eckhardt (2003) and Wedin \& Kerswell (2004). Just as for $E2\_0.625$,
subsequent investigations have revealed that $A3\_1.25$ is but one member
of another highly-symmetric class of TWs (the $M$-class, Pringle et
al. 2008) characterised by this distinctive double-layer structure.

After a time $200\, D/U$, the energy of the edge trajectory reaches
a constant plateau and the value of $r_{min}$ decreases
exponentially to $O(10^{-4})$ by $300 \, D/U$. This indicates that
the trajectory is converging to an end state. Newton-Krylov was used
to accelerate the convergence of the trajectory by taking the state
at $300\, D/U$ and reducing $r_{min}$ to $O(10^{-11})$.  The spatial
structure of the initial and final state of this procedure, however,
is indistinguishable. The final state is another travelling wave
solution, $C3\_1.25$ (see Figure \ref{plots2}), distinct from
$2b\_1.25$ of Kerswell \& Tutty (2007). Stability analysis of
$C3\_1.25$ reveals only one unstable direction which has to be
normal to the edge (see Pringle et al 2008 for details). Hence
$C3\_1.25$ is an attractor in the edge, and both $C3\_1.25$ and
$2b\_1.25$ can only be local attractors there. Closer examination of
$E2\_0.625$ and $C3\_1.25$, which look very similar, reveals that
they are members of the same TW family distinguished only by their
different wavelengths.

\subsection{Travelling waves on the Edge\label{sec:inedge}}

There is no guarantee that starting guesses taken from the edge for
the Newton-Krylov method will converge to states also on the edge.
The calculations of subsection 3.4 have, however, demonstrated this
for $A1\_0.625$ (the asymmetric TW) thereby confirming earlier
speculation (Schneider et al 2007b, Pringle \& Kerswell 2007). To
explicitly check this for the other TWs found here, each was taken
as an initial condition with various very small perturbations added
to start a series of runs. In all cases, two contrasting behaviours
were evident: one perturbation could be found which led to
relaminarisation while another led to a turbulent episode: see
Figure \ref{fig:edgechk}.  This confirms that all the TWs found here
are on the laminar/turbulent boundary for the particular parameter
settings tested (shown in Table 1).

\section{Discussion}\label{sec:disc}

A summary of the main findings of this investigation are listed below.\\
\begin{itemize}
\item The laminar-turbulent boundary $\Sigma$ in a short pipe seems
to have a chaotic attractor as observed  by Schneider et al.
(2007b).
\item Although $\Sigma$ has many unstable
lower branch TWs embedded within it (e.g. Kerswell \& Tutty 2007),
only  one - the  asymmetric TW of Pringle \&
Kerswell (2007) (and rotated derivatives of it) - was approached by
edge trajectories studied here.
\item A new asymmetric TW which has a small but finite azimuthal phase speed
has been discovered.
\item Some evidence was gathered for a ``relative" homoclinic connection
between the asymmetric TW and the same wave with a different
orientation.
\item Calculations within a $\R_2$-symmetric subspace have revealed
new branches of $\R_2$-symmetric TW solutions. One ($D2\_0.625$) is
the first TW found {\em not} to be shift-\&-reflect symmetric in
pipe flow, while the others are members of new classes of {\em
highly}\,-symmetric TWs (Pringle et al. 2008).
\item The laminar-turbulent boundary restricted to $\R_2$-symmetry has
at least two simple local attractors in the form of TWs at $Re=2400$.\\
\end{itemize}

\begin{figure}
\includegraphics[width=12cm]{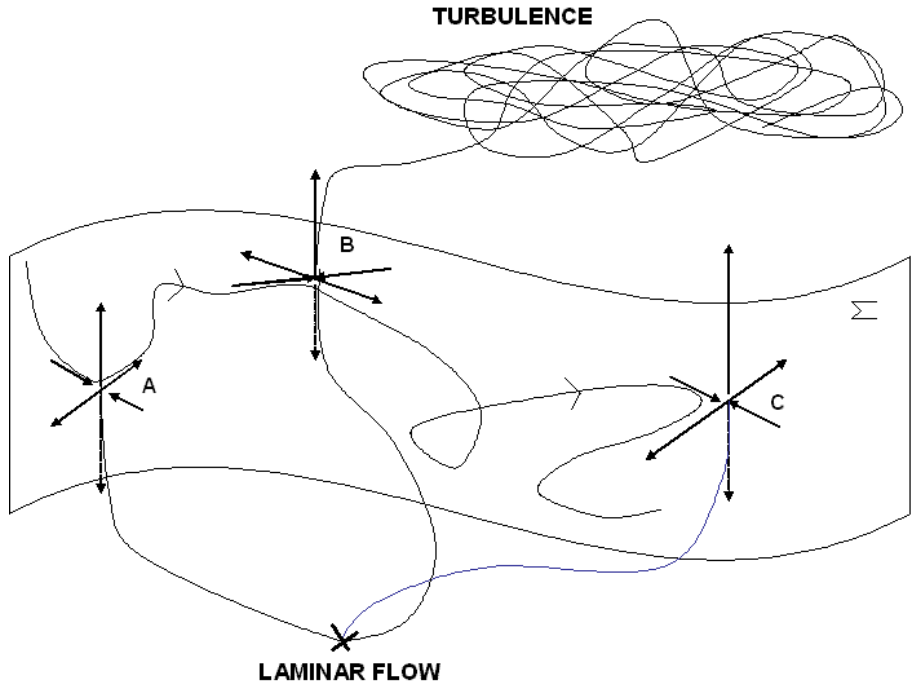}
\caption{Schematic view of phase-space. The surface $\Sigma$
separates initial conditions which relaminarise from those which
become turbulent. An edge trajectory visiting three states A, B and
C is shown schematically. The dynamics on the manifolds transverse
to $\Sigma$ are shown by trajectory diverging towards either the
laminar state or the turbulent state.}
\label{phasespace}
\end{figure}

\begin{figure}
 \begin{center}
 \setlength{\unitlength}{1cm}
  \begin{picture}(12,12)
   \put(0,5.2){\epsfig{figure=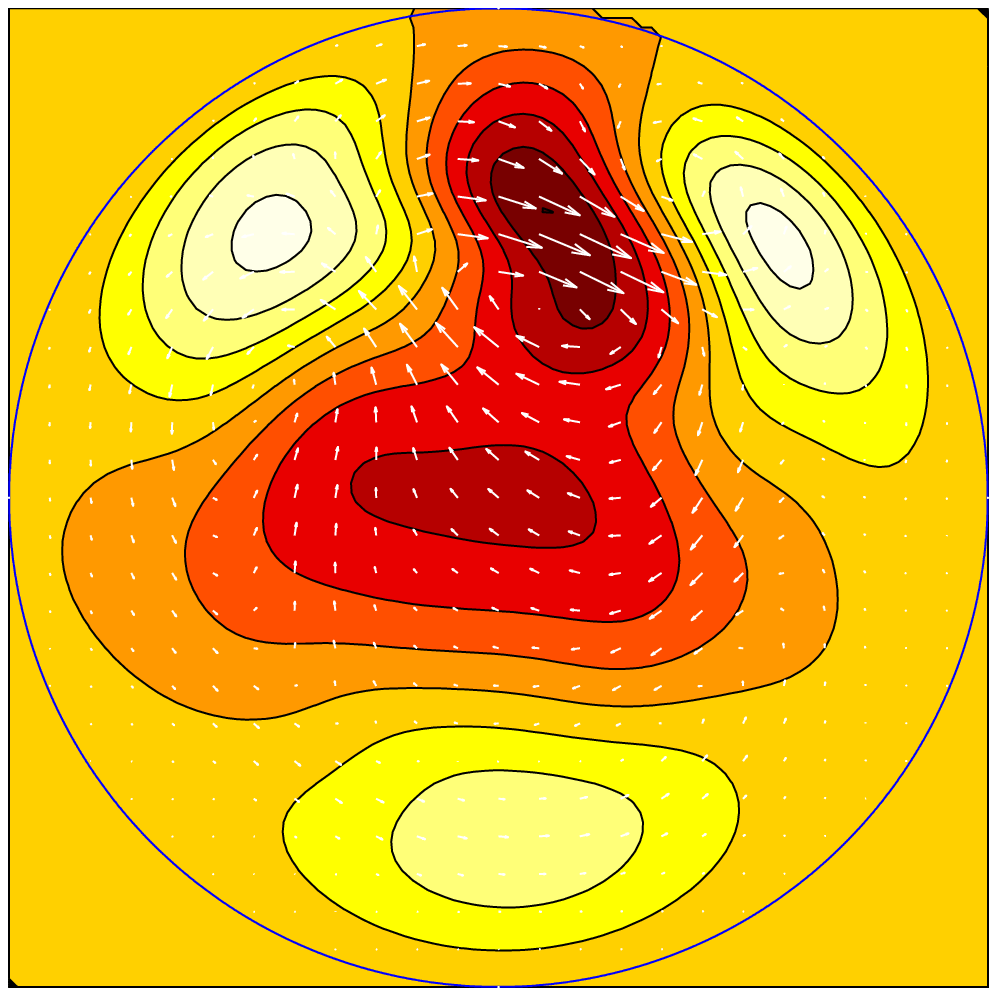,width=5cm,height=5cm,clip=true}}
   \put( 5.2,5.2){\epsfig{figure=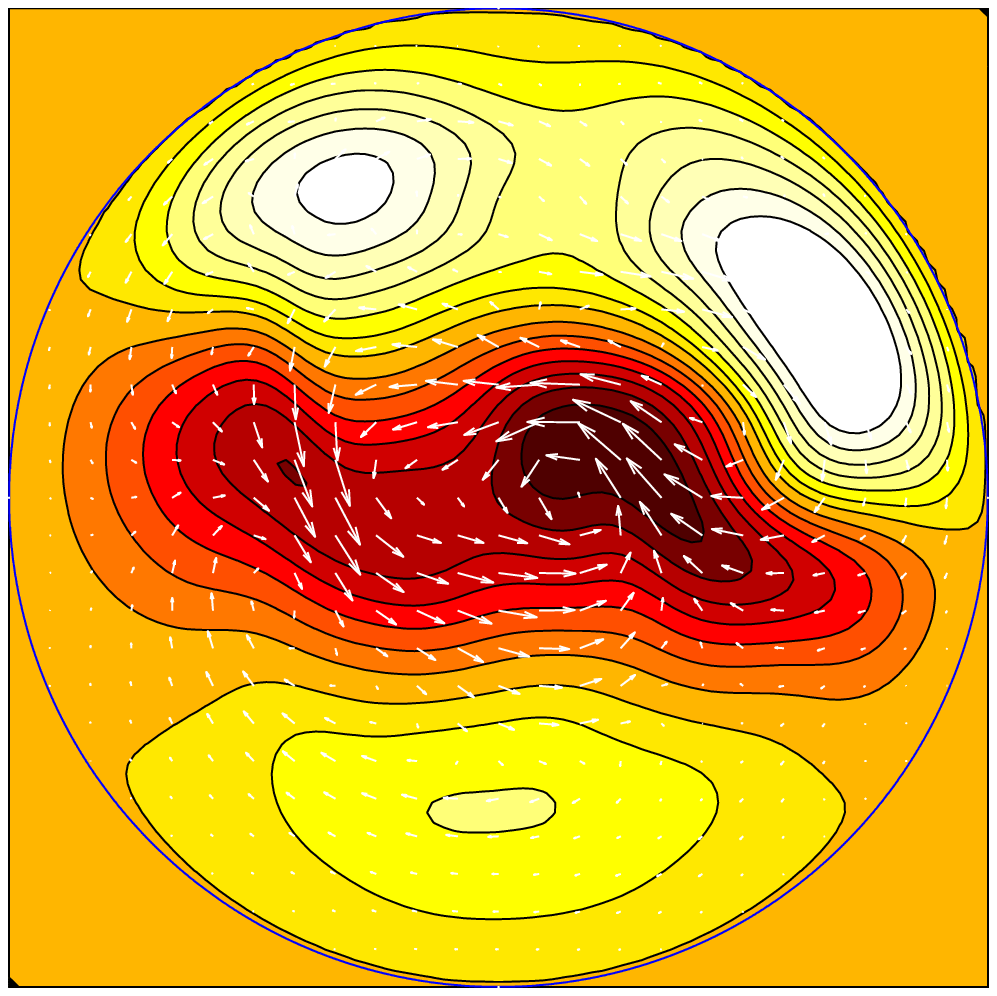,width=5cm,height=5cm,clip=true}}
   \put(0,0){\epsfig{figure=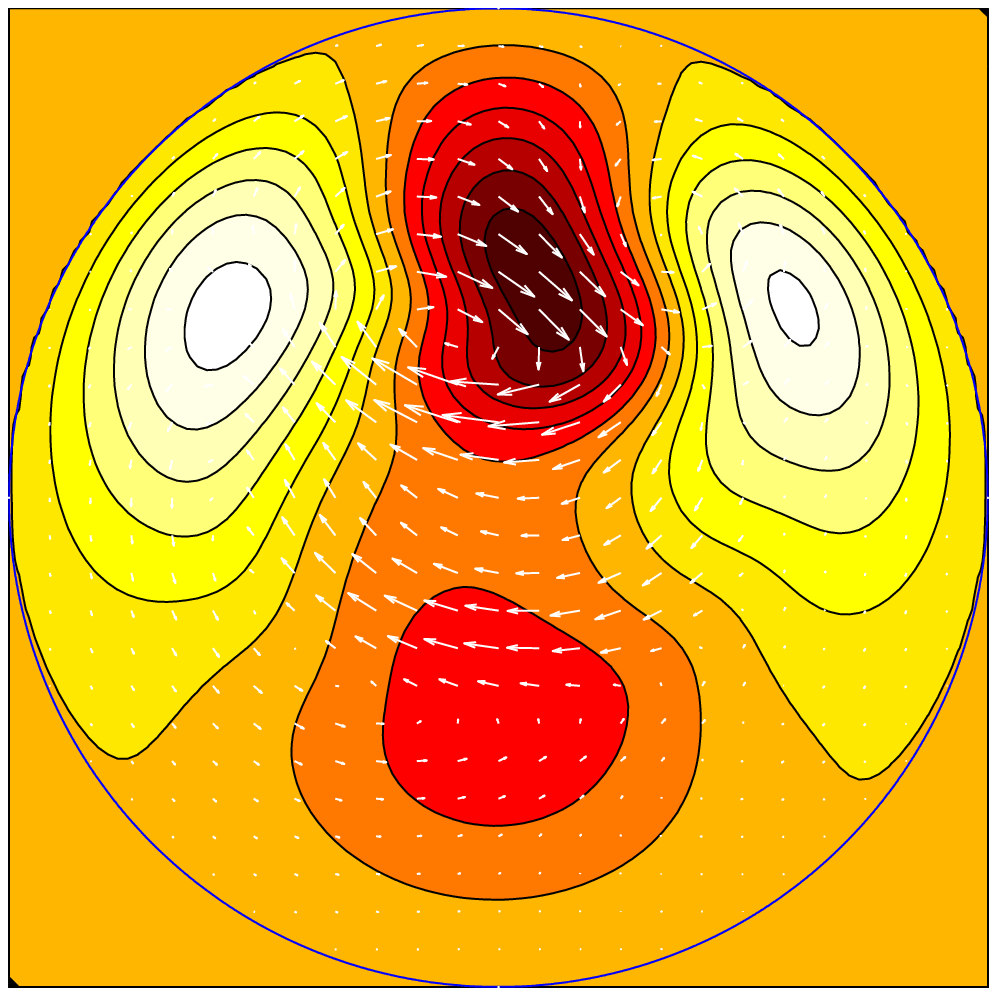,width=5cm,height=5cm,clip=true}}
   \put(5.2,0){\epsfig{figure=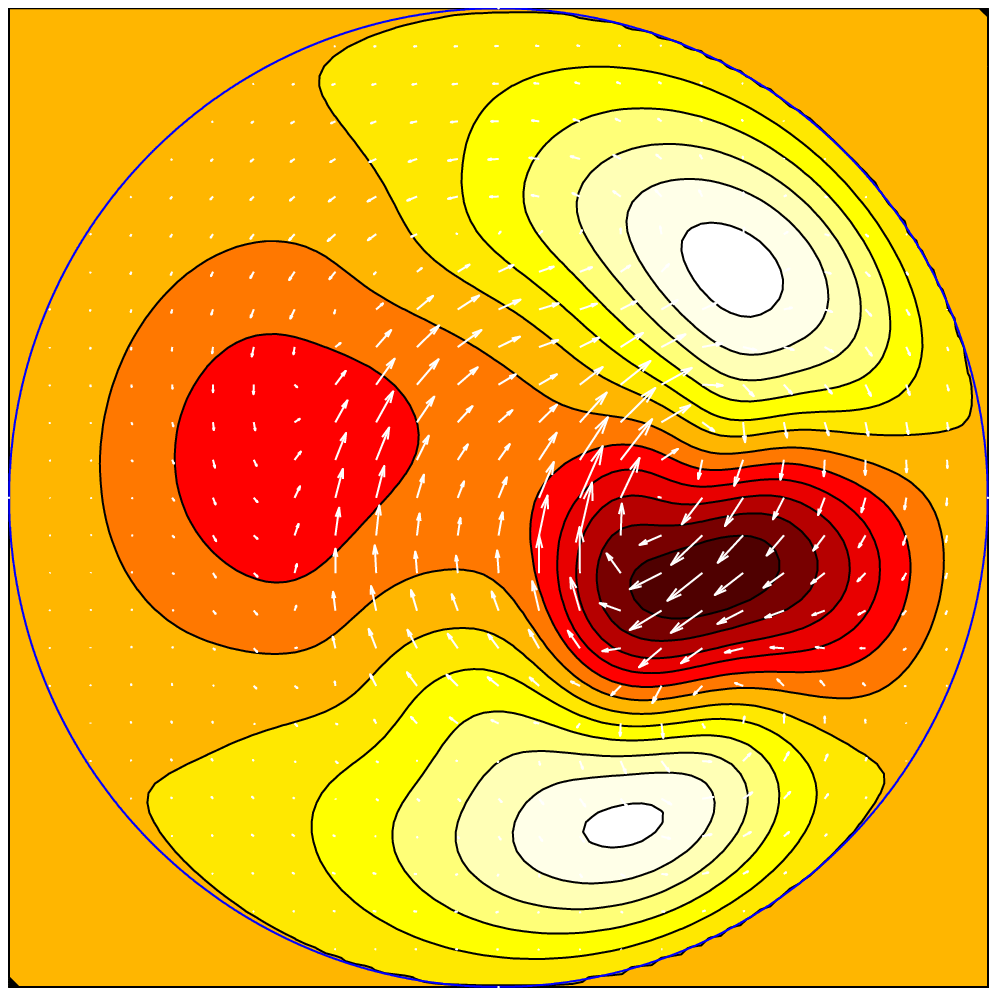,width=5cm,height=5cm,clip=true}}
   %\put(2,2){\int_0^1 sdx}
  \end{picture}
 \end{center}\caption{Starting guesses for the Newton-Krylov algorithm
 (top left $A1$, top right $B1$) and
 converged states (bottom left $A1\_0.625$, bottom right $B1\_0.625$)
 on the edge trajectory described in \S \ref{sec:rec}.
 Each subfigure represents a snapshot across the pipe. Contours indicate
 the axial velocity difference from the underlying laminar flow
 (light/dark indicating faster/slower moving fluid) and the arrows
 represent the cross-stream velocity (length proportional to speed).
 Maximum norm of the $(u,v)$ cross-velocity is $0.0128 U$ and the axial
 velocity differential $w$ is in the range $\pm 0.17 U$. The TWs
 $A1\_0.625$ and $B1\_0.625$
 are exactly the same state modulo an azimuthal shift.}
\label{plots1}
\end{figure}

\begin{figure}
\includegraphics[width=12cm]{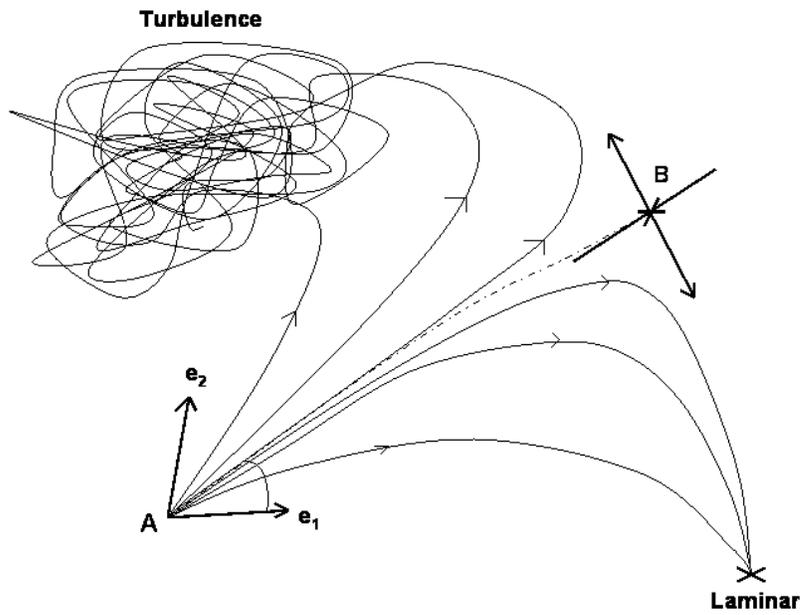}
\caption{A sketch of the heteroclinic connnection joining two saddle
points A and B. ${\bm e}_1$ and ${\bm e}_2$ are the two unstable
eigendirections of A, used for shooting by varying the shooting
angle, in order to optimise the approach to B.}
\label{heteroplot}
\end{figure}

\begin{figure}
\includegraphics{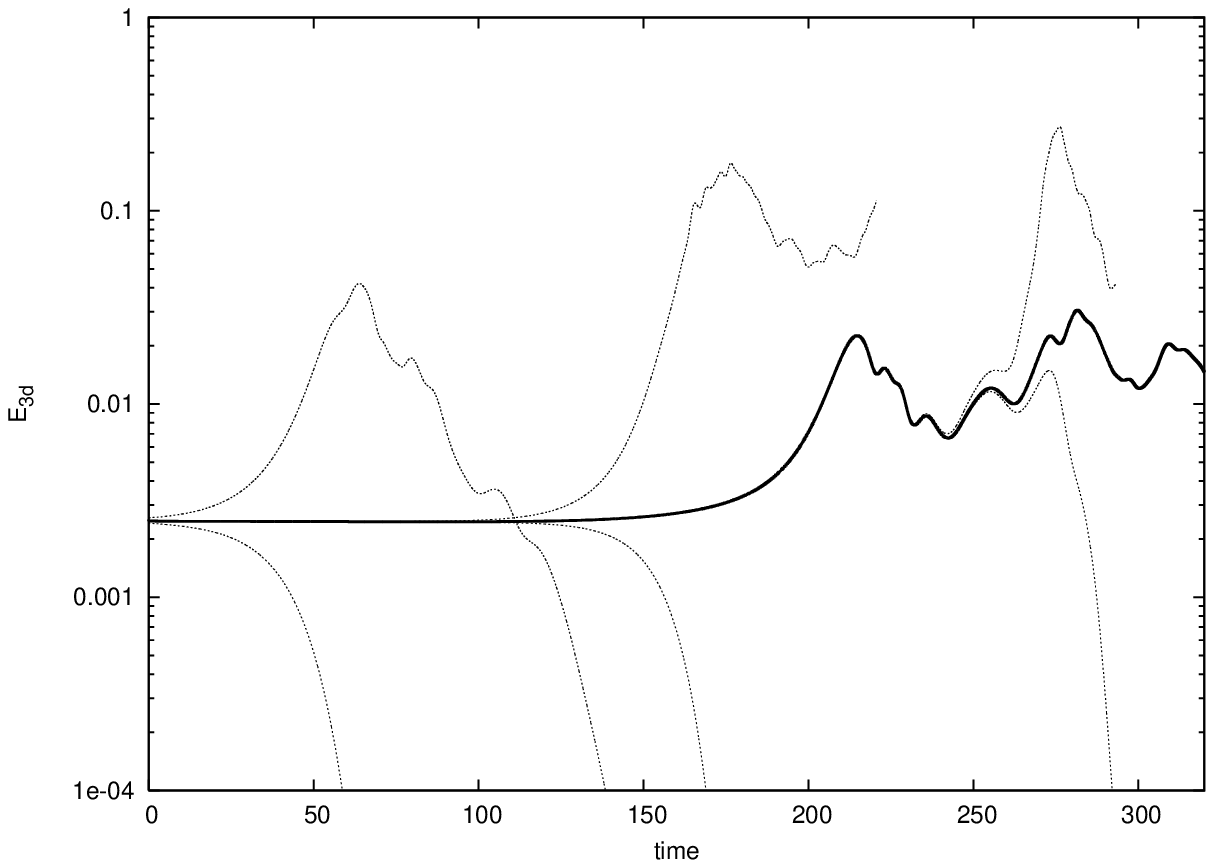}
\caption{Energy contained in the axially dependent modes for
$Re=2875$, $m_0=1$, $\alpha=0.625$. The thick line indicates the
edge trajectory and the thinner lines nearby trajectories which
either relaminarise (energy decreases) or becomes turbulent (energy
increases to a higher level). Time is in units of $D/U$}
\label{E4}
\end{figure}

\begin{figure}
\includegraphics{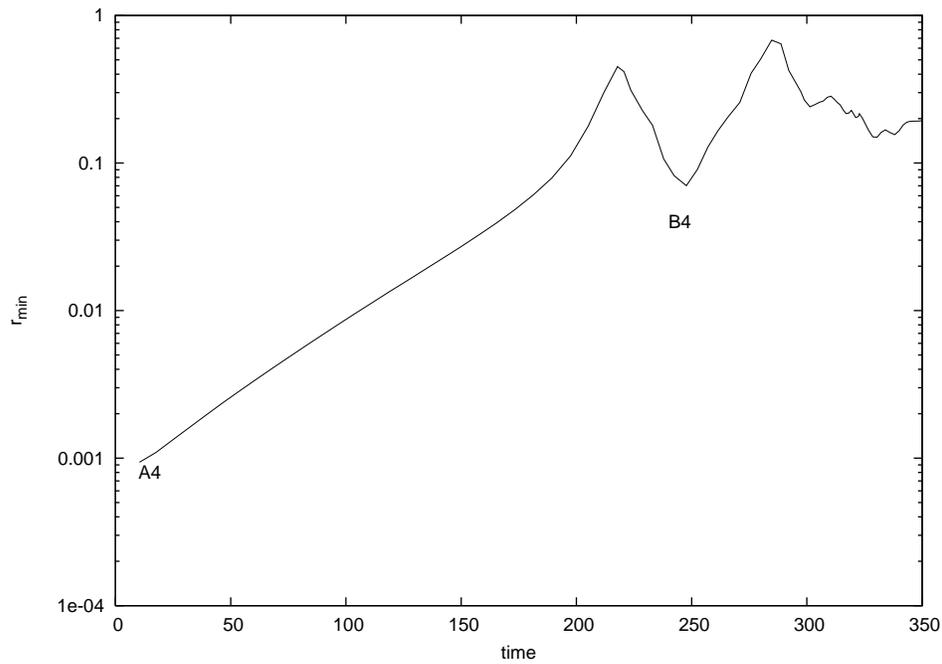}
\caption{Recurrence signal for the heteroclinic connection of \S
\ref{sec:heteroc} versus time (in $D/U$).
$(\alpha,Re,m_0)=(0.625,2875,1)$.}
\label{rmin4}
\end{figure}

\begin{figure}
\includegraphics{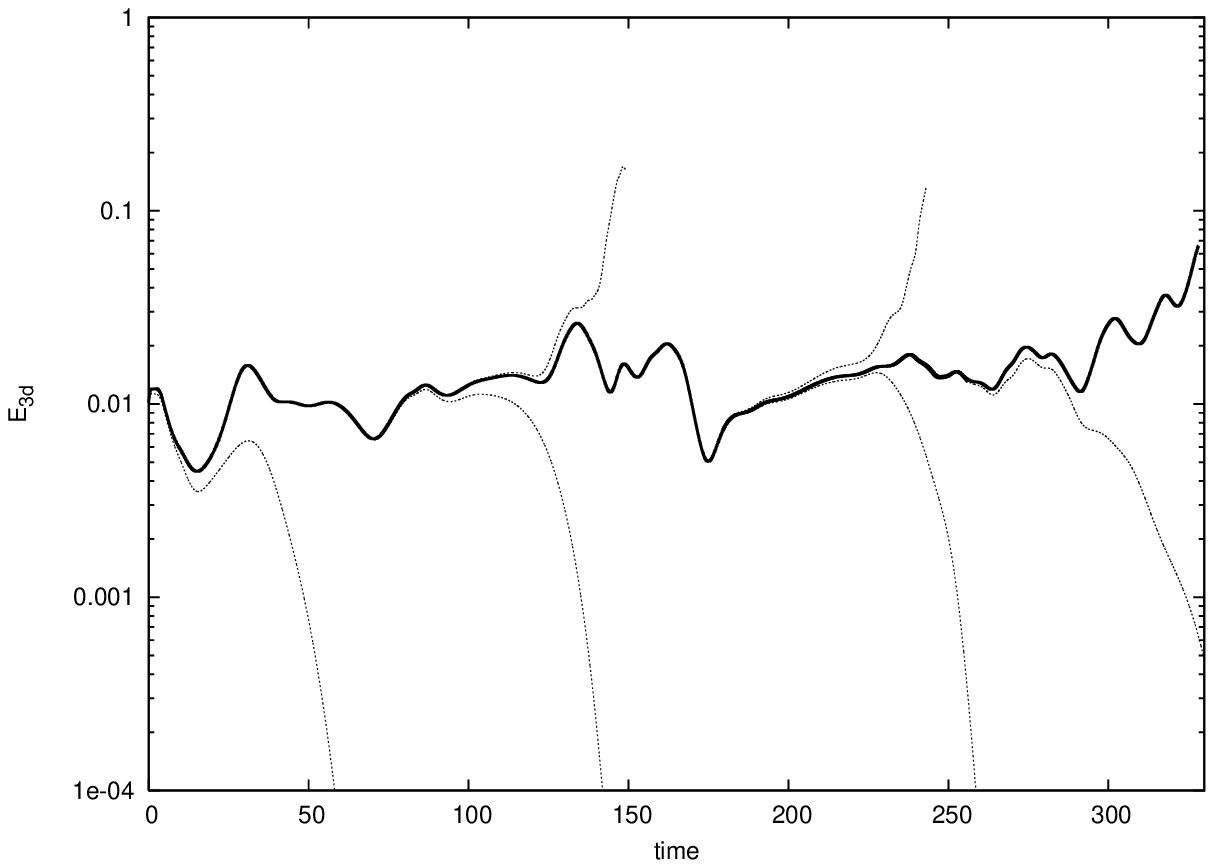}
\caption{Energy contained in the axially-dependent modes on the edge
for $(\alpha,Re,m_0)=(0.625,2875,2)$. The thick line indicates the
edge trajectory and the thinner lines nearby trajectories which
either relaminarise (energy decreases) or becomes turbulent (energy
increases to a higher level). Time is in units of $D/U$}
\label{E2}
\end{figure}

\begin{figure}
 \begin{center}
 \setlength{\unitlength}{1cm}
  \begin{picture}(12,12)
   \put(0,5.2){\epsfig{figure=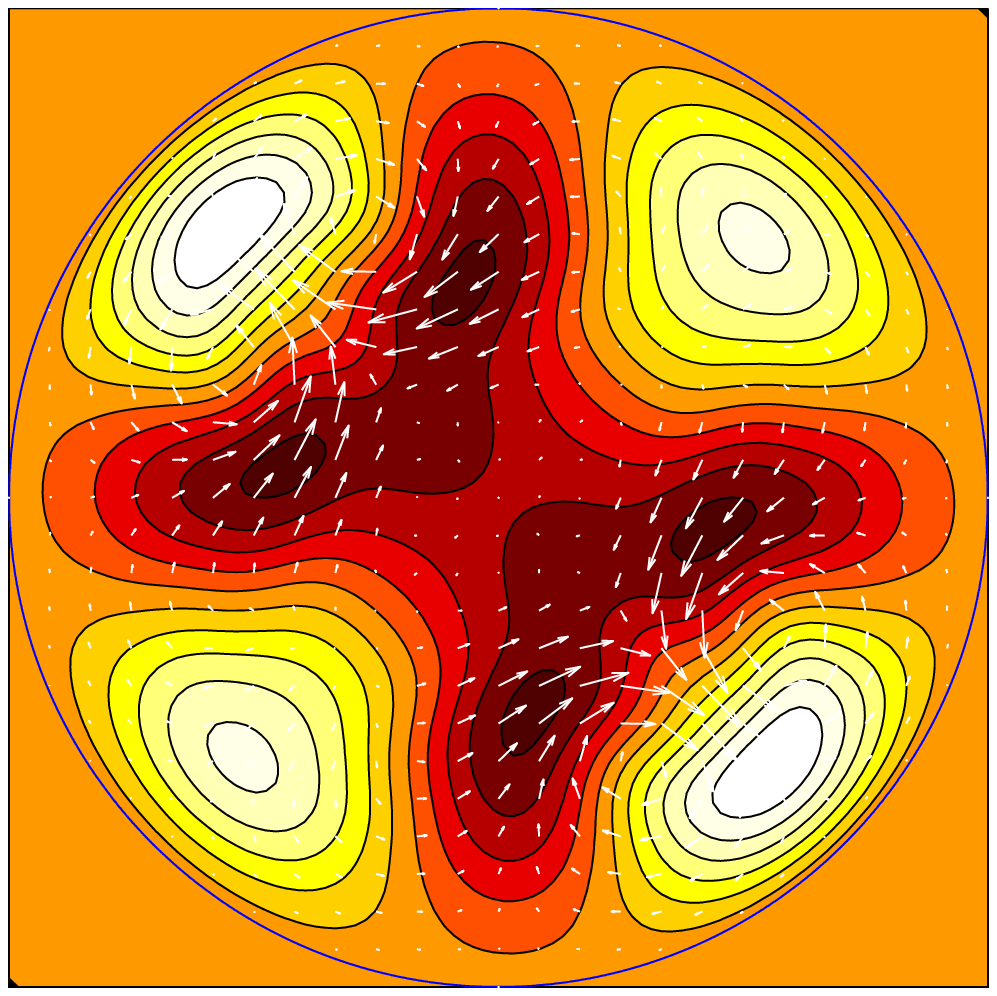,width=5cm,height=5cm,clip=true}}     \put(5.2,5.2){\epsfig{figure=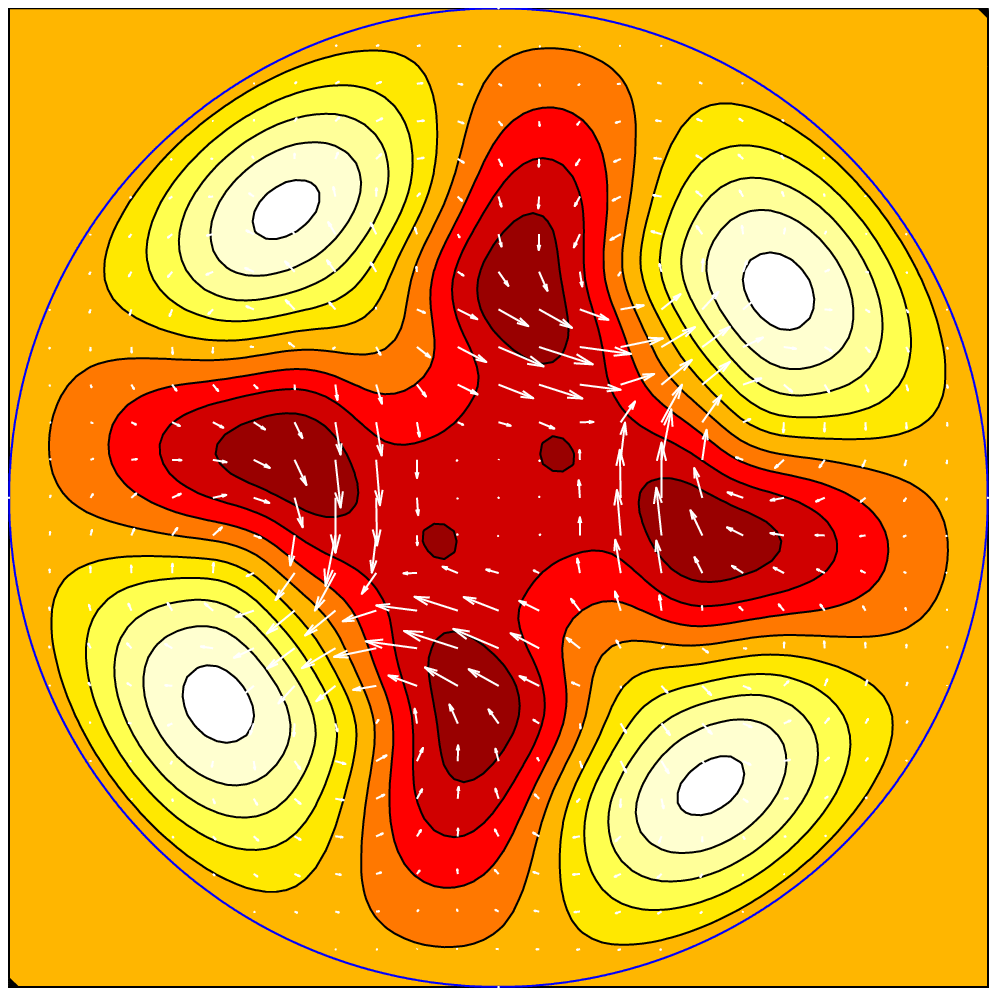,width=5cm,height=5cm,clip=true}}
  \put(0,0){\epsfig{figure=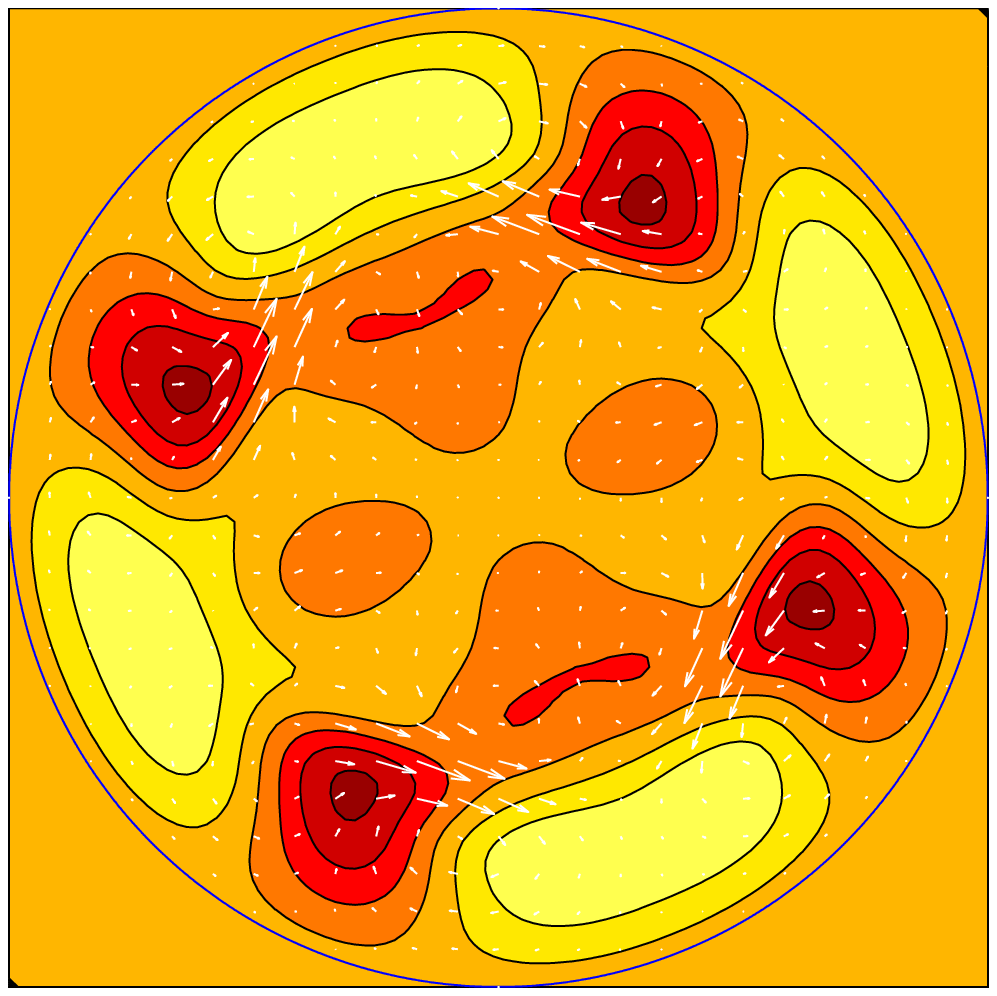,width=5cm,height=5cm,clip=true}}
  \put(5.2,0){\epsfig{figure=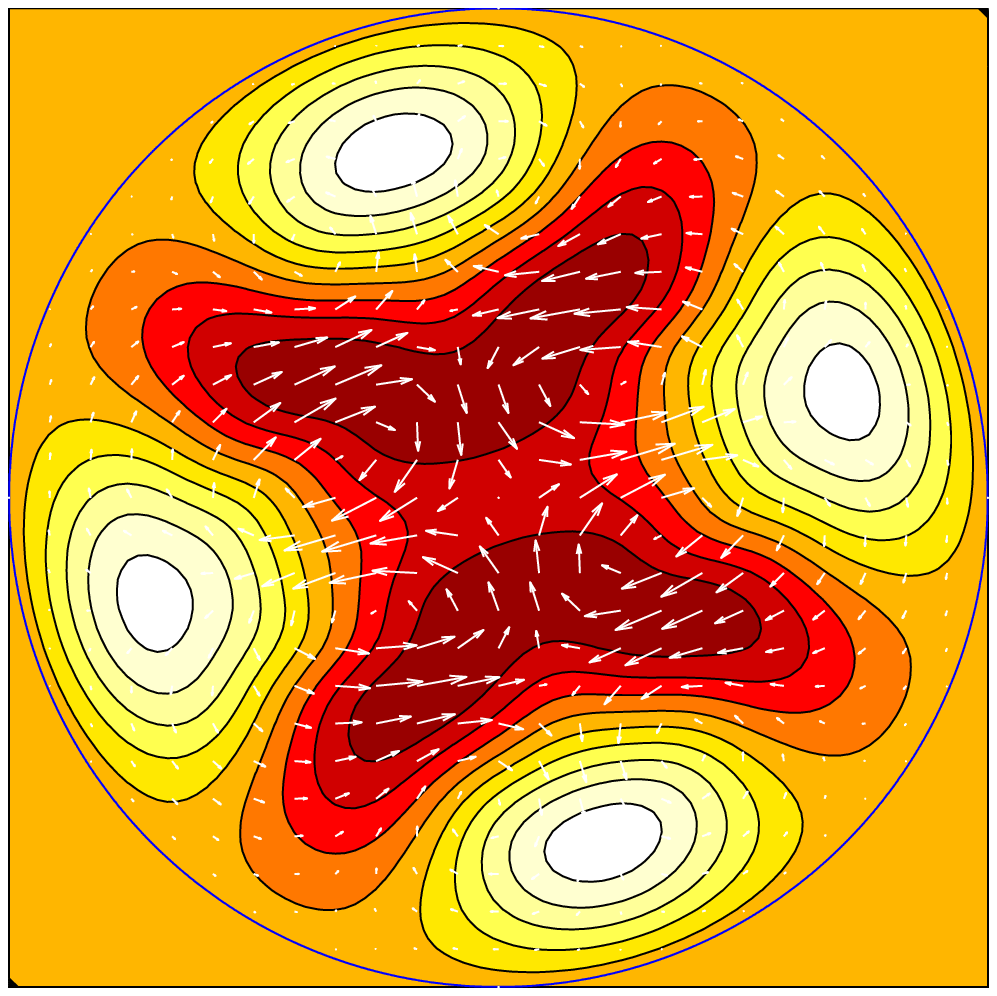,width=5cm,height=5cm,clip=true}}
  \end{picture}
 \end{center}\caption{Converged states $D2\_0.625$ (top, left),
 $E2\_0.625$ (top, right),
 $A3\_1.25$ (bottom, left) and $C3\_1.25$ (bottom,right).
 Contours indicate
 the axial velocity difference from the underlying laminar flow
 (light/dark indicating faster/slower moving fluid) and the arrows
 represent the cross-stream velocity (length proportional to speed).
 Across the four snapshots, the maximum cross-stream speed is $0.0143 U$,
 and the axial velocity perturbation $w$ is in the range $\pm 0.17 U$.}
\label{plots2}
\end{figure}

\begin{figure}
\includegraphics{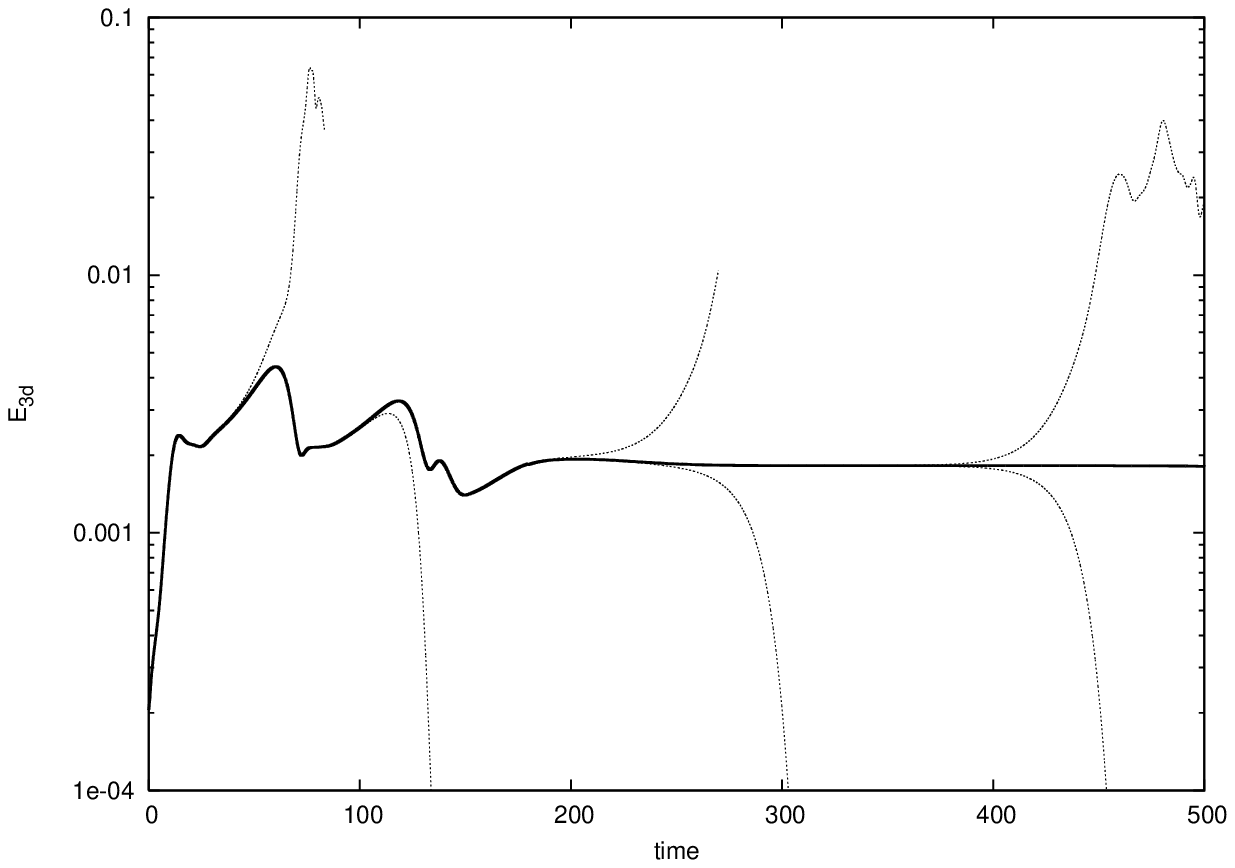}
\caption{Energy contained in the axially-dependent modes for
$(\alpha,Re,m_0)=(1.25,2400,2)$. The thick line indicates the edge
trajectory and the thinner lines nearby trajectories which either
relaminarise (energy decreases) or becomes turbulent (energy
increases to a higher level). Time is in units of $D/U$}
\label{E3}
\end{figure}

\begin{figure}
\includegraphics{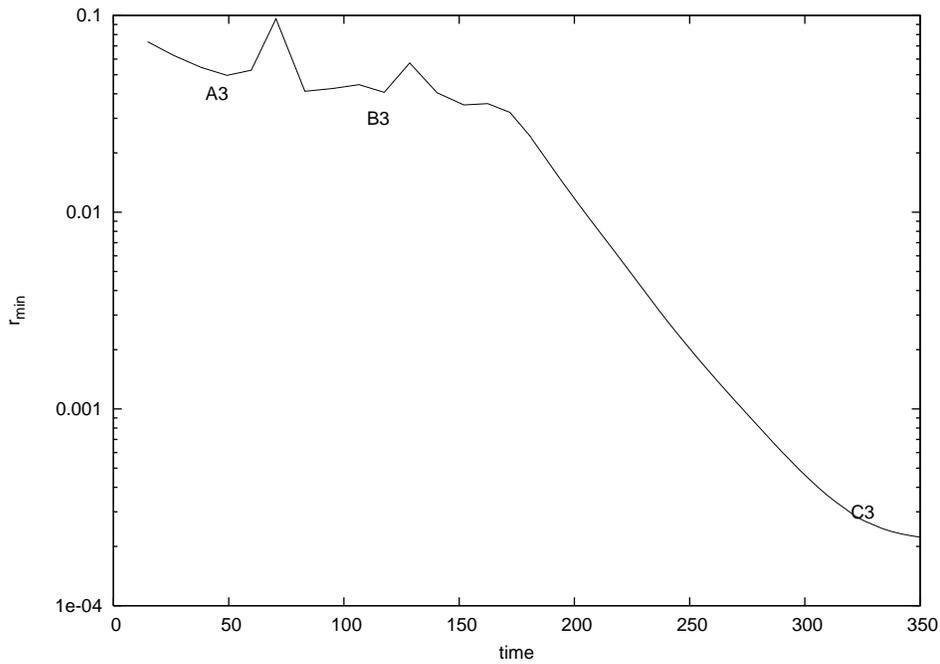}
\caption{Recurrence signal for the attracting state of Section
\ref{sec:m_02} versus time expressed in $D/U$ units.
$(\alpha,Re,m_0)=(1.25,2400,2)$.}
\label{rmin3}
\end{figure}

\begin{figure}
\includegraphics[width=12cm]{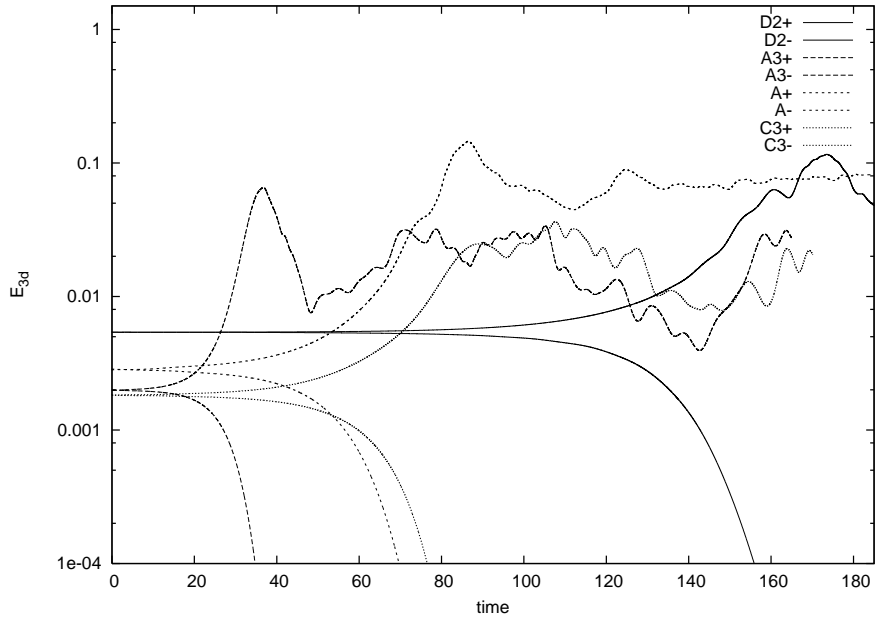}
\caption{
  Diverging trajectories demonstrating the `edge credentials' of the
  new TWs found here. Two typical trajectories starting close to each
  TW are shown with one decaying down smoothly to the laminar state
  ($E_{3d}=0$) and another increasing up to the turbulent state. $D2$
  represents $D2\_0.625$, $A3$ is $A3\_1.25$, $A$ is $A1\_0.625$ and $C3$
  is $C3\_1.25$ with $\pm$ used to distinguish the starting
  perturbation.}
\label{fig:edgechk}
\end{figure}

Our numerical experiments have shed some light on the dynamical
structure of $\Sigma$ in pipe flow. When no symmetry is imposed, and
the pipe is approximately $5\,D$ long ($\alpha=0.625$), trajectories
which neither relaminarise nor become turbulent all look chaotic and
visit some exact recurrent states in a transient manner. In all cases
studied, these exact solutions appear to be specifically the
asymmetric TWs found by Pringle \& Kerswell (2007) even though many
other lower branch TWs are embedded in $\Sigma$ (Kerswell \& Tutty
2007). No RPOs were found even though the numerical method was
sufficiently general to find them, in addition to TWs. Whether this is
because the RPOs are rarely visited or just more difficult to isolate
is unclear. It is tempting to conclude that chaos on the edge of pipe
flow is structured around a set of unstable saddle points (the TWs)
linked together by heteroclinic or sometimes relative heteroclinic
connections. An approximation to such a connection has been shown in
subsection \ref{sec:heteroc}. The resulting dynamical structure is
likely to be a heteroclinic (or homoclinic) tangle, an efficient
mechanism to produce chaotic trajectories in phase-space with the set
of TWs acting as a skeleton of the tangle.  Seen from this point of
view, transition to turbulence from a given initial condition depends
on the position of the initial state, in phase space, relative to the
stable manifold of each exact state embedded in $\Sigma$. When a
trajectory approaches a travelling wave solution belonging to
$\Sigma$, its relative position to the boundary determines on which
side of the edge the trajectory escapes, resulting in either
relaminarisation or a turbulent transient (see sketch in Figure \ref{phasespace}). \\

\begin{table}
\begin{center}
\begin{tabular}{@{}ccccccc@{}}
 TW  & $m_0$ & $Re$ & $C$ & $c_{\theta}$ & status\\
  \hline
  & & & & & \\
  $A1\_0.625$ & 1 & 2875 &  1.55494 & 0                & known\\
  $A'\_0.625$ & 1 & 2875 &  1.51956 & $-1.704.10^{-3}$ & new\\
  & & & & & \\
  $D2\_0.625$ & 2 & 2875 &  1.53382 & 0                & new \\
%  $E2\_0.625$ & 2 & 2875 &  1.57350 & 0                & new \\
  $A3\_1.25$  & 2 & 2400 &  1.23818 & 0                & new\\
  $C3\_1.25$  & 2 & 2400 &  1.55064 & 0                & new\\
  & & & & & \\
\end{tabular}
\end{center}
  \label{TWtable}
  \caption{
  Parameters, axial propagation speed $C$ (in units of $U$)
  and azimuthal propagation speed $c_{\theta}$ in $rad\,U/D$ and
  status  of all travelling wave solutions found in this paper
  ($E2\_0.625$ is in the same TW family as $C3\_1.25$ and hence not listed separately).
  }
\end{table}

When the flow is artificially restricted to be $\R_2$-symmetric at
$Re=2875$, the trajectory remains chaotic wandering sufficiently
close to \emph{distinct} TW states ($D2\_0.625$ and $E2\_0.625$) to
be captured by the Newton-Krylov scheme. By halving the pipe length
to $L \approx 2.5\,D$ and reducing $Re$ to $2400$, multiple local
attractors appear in $\Sigma$. A randomly-started trajectory, after
a chaotic transient and approaches to states $A3\_1.25$ and
$B3\_1.25$, is attracted towards the TW solution $C3\_1.25$. The
fundamental difference between $C3\_1.25$ and all the other states
mentioned so far lies in the number of unstable eigenvalues. When a
TW has two or more unstable eigenvalues in a given subspace, the
dimension of the intersection between its unstable manifold and the
hypersurface $\Sigma$ is reduced by one but remains at least one.
Hence such a state remains a saddle point on $\Sigma$: edge
trajectories enter its vicinity along its stable manifold and escape
along the unstable manifold. In contrast, when a state has only one
unstable direction, this is necessarily normal to $\Sigma$, so that
the state becomes an attractor for dynamics restricted to the edge
(a relative attractor). We have shown that $C3\_1.25$ is one such
TW, but there is at least one other, $2b\_1.25$ of Kerswell \& Tutty
(2007). Therefore we expect both these two states to be only
\emph{local} attractors on $\Sigma$ rather than global ones. In the
cases of plane Poiseuille and Couette flow, a lower-branch solution
exists which has only one unstable direction and is therefore a
relative attractor (Toh \& Itano 1999, Itano \& Toh 2001, Wang {\it
et. al.} 2007, Schneider et al., 2008, Viswanath 2008). Pipe flow is
different as TWs all seem to have at least two unstable directions
when there is no imposed discrete rotational symmetry. For example,
the $2b\_1.25$ TW has 1 harmonic ($\R_2$-symmetric) unstable
eigenfunction and one subharmonic (only $\R_1$-symmetric)
unstable eigenfunction.\\

While discrete rotational symmetry constraints might appear
artificial, they have nevertheless served as a useful device for
discovering new exact recurrent solutions (see Table 1).  The method
developed here based on edge tracking, recurrence analysis and use
of a Newton-Krylov algorithm, is a general approach for finding new
exact recurrent solutions in any flow situation possessing
subcritical behaviour. Importantly, the present method naturally
selects the states that are most likely to be visited and does not
presuppose anything about their spatial structure (modulo the
symmetries imposed on the flow). The use of the scalar function
$r_{min}$, coupled with a Newton-Krylov solver, can also be used  to
search for more complex solutions such as relative periodic orbits,
whether located on the laminar-turbulent boundary or embedded in
fully developed turbulence. This is currently underway.\\

May open questions remain regarding $\Sigma$ in pipe flow. Is
$\Sigma$ really a hypersurface or can it have a more fractal
structure? Why does the flow keep approaching the asymmetric TW or
rotations of it, and not any of the other lower branch TWs known to
exist within the same parameters range? Finally, what are the
properties of $\Sigma$ for longer pipes, where localised turbulent
`puffs' structures exist? Preliminary work for the extended pipe in
a reduced model has already revealed an interesting localised
structure as the attracting state (Willis \& Kerswell 2008b).\\

\vspace{0.5cm} \noindent Acknowledgements:

\begin{acknowledgments}
We would like to thank C. Pringle for helping with the velocity
profiles of the TW solutions and B. Eckhardt for stimulating
discussions about this topic. Y.D. was supported by a Marie-Curie
Intra-European Fellowship (grant number MEIF-CT-2006-024627) and A.W.
by the EPSRC (grant number GR/S76144/01).

\end{acknowledgments}

%***************************************************************

\

%--------------------------------------------------------------------------

% fig 1

% fig 2

% fig 3

% fig 4

% fig 5

% fig 6

% fig 7

% fig 8

% fig 9

% fig 10

% fig 11

% fig 12

% fig 13

% fig 14

%--------------------------------------------------------------------------
\end{document}